# Calibration of the Sunspot and Group Numbers Using the Waldmeier Effect

Leif Svalgaard[1] (leif@leif.org), David H. Hathaway[2]

[1] Stanford University, Cypress Hall C3, W.W. Hansen Experimental Physics Laboratory, Stanford University, Stanford, CA 94305, USA

[2] (Retired), NASA Ames Research Center, Moffett Field, CA 94035, USA

**Abstract:**

The Waldmeier Effect is the observation that the rise time of a sunspot cycle varies inversely with the cycle amplitude: strong cycles rise to their maximum faster than weak cycles. The shape of the cycle and thus the rise time does not depend on the scale factor of the sunspot number and can thus be used to verify the constancy of the scale factor with time as already noted by Wolfer (1902) and Waldmeier (1978). We extend their analysis until the present using the new SILSO sunspot number (version 2) and group number and confirm that the scale factors have not varied significantly the past 250 years. The effect is also found in sunspot areas, in an EUV (and F10.7) proxy (the daily range of a geomagnetic variation), and in Cosmic Ray Modulation. The result is that solar activity reached similar high values in every one of the (17$^{th}$?) 18$^{th}$, 19$^{th}$, and 20$^{th}$ centuries, supporting the finding that there has been no modern *Grand* Maximum.

# 1. Introduction

Max Waldmeier (1978) reminded us that „Es besteht eine Relation zwischen der Aufsteigzeit *T* (in Jahren) vom Minimum zum Maximum und der gröβten ausgeglichenen monatliche Relativzahl *RM*: log *RM* = 2.73 - 0.18 *T*. Die Epochen der Extrema lassen sich bestimmen ohne Kenntnis der Reduktionsfactoren. Da die genannte Beziehung zwischen *T* und *RM* auch für dies Zeit von 1750 bis 1848 gültig ist, besteht Gewähr, daβ der Skalenwert der Relativzahlen seit über 200 Jahren konstant geblieben ist oder nur unbedautenden Schwankungen unterworfen ist.[1]" Figure 1 shows how to first order, the sunspot numbers for different solar cycle form a one-parameter set of curves, the shapes of which are determined by the maxima of the (smoothed) values of the sunspot number, in that larger cycles reach their maximum earlier in the cycle than smaller cycles.

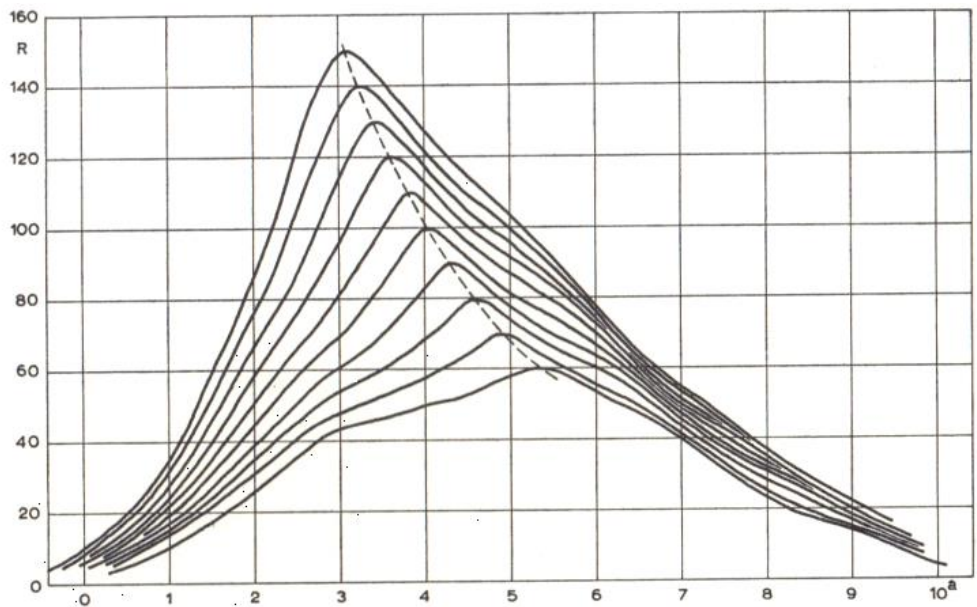

Figure 1: Waldmeier's one-parameter curve family for the sunspot number through the cycle (Waldmeier 1968).

As Waldmeier pointed out, the shape of the sunspot cycle curve, and thus the rise time from minimum to maximum, do not depend on the 'scale value' of the sunspot number. Determination of the rise time can therefore be used to verify if the scale value has changed in the past. His conclusion was that it had not changed for more than the last 200 years. The 'new' sunspot number (Clette et al. 2014) series differs from the old Zürich series mainly by removing the artificial 0.6 factor introduced by Wolfer to reduce his counts to the scale of Wolf's. This simply adds log(1/0.6) = 0.22 to the Waldmeier relation to read log *RM* = 2.95 – 0.18 *T*, and does not change Waldmeier's conclusion that the scale value has not varied significantly or systematically since 1750. This conclusion is backed by the reconciliation of the Sunspot Group Number (Svalgaard and Schatten, 2015) with the new sunspot number (Clette and Lefèvre, 2016).

---

[1] There is a relationship between the rise time *T* (in years) from minimum to maximum and the maximum smoothed monthly sunspot number *RM*: log *RM* = 2.73 – 0.18 *T*. The times of the extrema can be determined without knowledge of the reduction (or scale) factors. Since this relationship also holds for the years from 1750 to 1848 we can be assured that the scale value of the relative sunspot number over the last more than 200 years has stayed constant or has only been subject to insignificant variations.

## 2. Waldmeier Effect, Revisited

Figure 2 shows the two largest pairs of solar cycles in each of the three last full centuries in a modern setting using the revised Sunspot Number (SN) and Group Number (GN) scaled by a factor of 20 to match the SN (http://www.sidc.be/silso/datafiles).

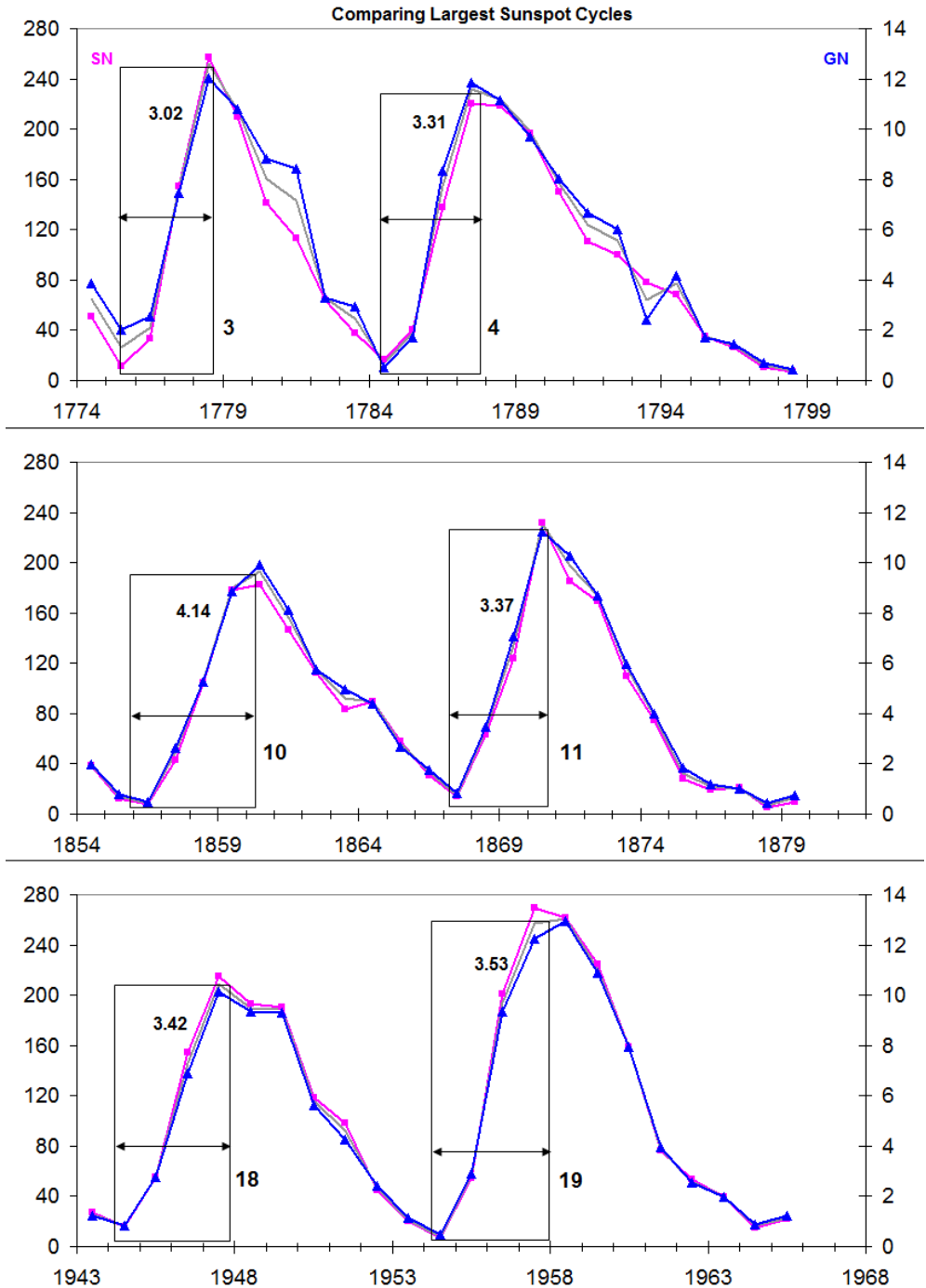

Figure 2: The largest cycles (numbers 3 and 4 in the 18th, 10 and 11 in the 19th, and 18 and 19 in the 20th century) as given by the revised sunspot (SN pink curves) and group numbers (GN blue curves) scaled by a factor of 20 (=12.08/0.6 which evidently yields a close match to the SN). The grey curve shows the average of SN and 20 times GN. The boxes span the rise from minimum to maximum annotated with the rise time in years taken from Table 1.

The cycles are of about equal amplitude (mean SN 228) and rise time (3.46 years) showing that solar activity rose to the same level in each century. Waldmeier (1958) pointed out that this was the case not only for yearly averages but also for the extreme values (old scale, version 1): „Auch die sehr seltenen Tage mit R > 300 haben sich im laufenden Zyklus gehäuft, indem 1956 zwei und 1957 elf solche auftreten sind. Während des ebenfalls intensiven Zyklus Nr. 18 sind nur vier solche Tage verzeichnet worden (1947/48). Wir müssen bis ins Jahr 1870 zurückgehen, um weitere Relativzahlen > 300 anzutreffen. Damals wurde diese Grenze am 26., 27. und 29. August mit 317, 312 bzw. 311 überschritten. Schliesslich stieg die Relativzahl in Jahre 1778 (wo die Beobachtungen sehr Lückenhaft sind) an neun Tagen auf über 300."[2]

Figure 3 shows the smallest sunspot cycles for the last three full centuries:

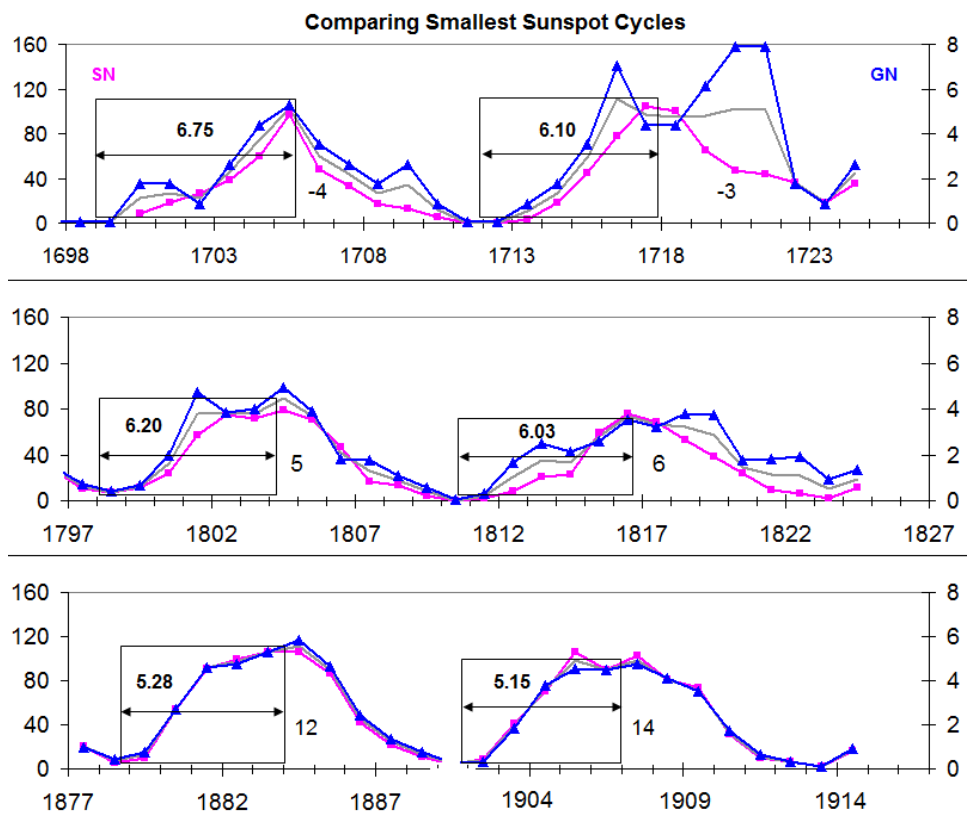

Figure 3: The smallest sunspot cycles since year 1700 in the same format as Figure 2.

---

[2] Also, the very rare days with R > 300 have been abundant in the current cycle, in that in 1956 there have been two and in 1957 eleven such days. During the also intensive Cycle 18 only four such days were registered. We have to go back to the year 1870 to come across Relative Sunspot Numbers > 300. Back then, this limit was exceeded on the 26th, 27th, and 29th August with 317, 312, and 311, respectively. Finally, the relative number in 1778 (even though the coverage was very poor) climbed to above 300 on nine days.

These small cycles are also of about equal amplitude (mean SN 92) and rise time (5.92 years). Figure 3 also illustrates the difficulty of defining the rise time and the maximum sunspot number, especially for small cycles. Usually, some kind of 'smoothing' or suppression of short-period variations is employed to define 'extrema' of the cycle, e.g. Meeus (1958) and similar ad-hoc methods. The traditional 13-month running mean (centered on a given month with equal weights for months –5 to +5 and half weight for months –6 and +6) is widely used but does a poor job of filtering out high frequency variations, although it is better than the simple yearly average. We discuss filtering in Section 3. Here we continue with estimates of the Waldmeier Effect culled from the literature. Table 1 contains a selection of rise times arrived at in different ways.

Table 1: Rise Times (T years) and Sizes (SN) of Sunspot Cycles -4 through 24 as reported in the literature. T1 is from Wolf, Brunner (1939), and Waldmeier (1961, 1978), T2 is from Chumak (2010), T3 from Kane (2008), T4 from Tritakis (1982), T5 from visual "eyeballing" (Svalgaard, this article), T6 (12-month running mean) and T7 (24-month Gaussian filtering) are from Hathaway (2015). The rightmost column contains log(max SN) calculated using Waldmeier's formula.

| Cycle | T 1 | T 2 | T 3 | T 4 | T 5 | T 6 | T 7 | Avg T | SN | log SN | log SN' |
|---|---|---|---|---|---|---|---|---|---|---|---|
| -4 | 7.5 | | | | 6.0 | | | **6.75** | 80 | **1.903** | 1.735 |
| -3 | 6.2 | | | | 6.0 | | | **6.10** | 105 | **2.021** | 1.852 |
| -2 | 4.0 | | | | 5.0 | | | **4.50** | 200 | **2.301** | 2.140 |
| -1 | 4.7 | | | | 5.0 | | | *4.85* | *150* | *2.176* | 2.077 |
| 0 | 5.3 | | | | 5.0 | | | *5.15* | *155* | *2.190* | 2.023 |
| 1 | 6.3 | | 6.25 | | 5.0 | 6.33 | 6.25 | **5.95** | 150 | **2.176** | 1.879 |
| 2 | 3.2 | | 3.42 | | 3.0 | 3.25 | 3.58 | **3.30** | 195 | **2.290** | 2.356 |
| 3 | 2.9 | | 2.92 | | 3.0 | 2.92 | 3.25 | **3.02** | 249 | **2.396** | 2.407 |
| 4 | 3.4 | | 3.33 | | 3.0 | 3.42 | 3.50 | **3.31** | 228 | **2.358** | 2.355 |
| 5 | 6.9 | | 6.75 | | 5.0 | 6.83 | 6.17 | **6.20** | 86 | **1.934** | 1.833 |
| 6 | 5.8 | | 5.33 | | 7.0 | 5.75 | 6.00 | **6.03** | 73 | **1.863** | 1.864 |
| 7 | 6.6 | | 6.42 | | 6.0 | 6.50 | 6.42 | **6.36** | 122 | **2.086** | 1.806 |
| 8 | 3.3 | | 3.33 | 3.5 | 4.0 | 3.33 | 3.42 | **3.51** | 227 | **2.356** | 2.318 |
| 9 | 4.6 | | 4.08 | 4.7 | 4.5 | 4.58 | 4.92 | **4.56** | 200 | **2.301** | 2.129 |
| 10 | 4.1 | | 4.17 | 4.2 | 4.0 | 4.17 | 4.25 | **4.14** | 190 | **2.279** | 2.204 |
| 11 | 3.4 | | 3.50 | 3.3 | 3.0 | 3.42 | 3.67 | **3.37** | 228 | **2.358** | 2.343 |
| 12 | 5.0 | | 5.17 | 5.3 | 6.0 | 5.00 | 4.92 | **5.28** | 110 | **2.041** | 2.000 |
| 13 | 4.5 | | 3.92 | 4.6 | 4.0 | 3.83 | 3.50 | **4.10** | 148 | **2.170** | 2.211 |
| 14 | *5.3* | | 5.33 | 5.8 | 5.0 | 4.08 | 4.33 | **5.15** | 97 | **1.987** | 2.022 |
| 15 | 4.0 | | 3.92 | 4.2 | 4.0 | 4.08 | 4.42 | **4.11** | 165 | **2.217** | 2.211 |
| 16 | 4.8 | | 4.67 | 4.9 | 4.5 | 4.67 | 4.33 | **4.64** | 125 | **2.097** | 2.115 |
| 17 | 3.6 | | 3.67 | 3.7 | 4.0 | 3.58 | 4.17 | **3.83** | 189 | **2.276** | 2.261 |
| 18 | 3.3 | | 3.25 | 3.2 | 3.5 | 3.25 | 3.83 | **3.42** | 209 | **2.320** | 2.335 |
| 19 | 3.6 | 3.6 | 3.83 | 3.3 | 3.0 | 3.92 | 3.83 | **3.53** | 260 | **2.415** | 2.315 |
| 20 | 4.2 | 4.0 | 4.00 | 4.2 | 5.0 | 4.08 | 4.42 | **4.30** | 150 | **2.176** | 2.176 |
| 21 | 3.4 | 3.4 | 3.75 | | 3.5 | 3.75 | 4.17 | **3.64** | 224 | **2.350** | 2.294 |
| 22 | | 2.8 | 3.08 | | 4.0 | 2.83 | 3.42 | **3.33** | 205 | **2.312** | 2.352 |
| 23 | | 4.0 | 3.92 | | 5.0 | 3.92 | 4.58 | **4.37** | 176 | **2.246** | 2.163 |
| 24 | | | | | 5.5 | 5.42 | 5.08 | **5.29** | 123 | **2.090** | 1.998 |

Using the data from Table 1 (with the new sunspot number series from WDC/SILSO) we can now revisit the Waldmeier Effect, Figure 4, and we confirm that Waldmeier's conclusion about the stability of the sunspot number 'scale value' still holds.

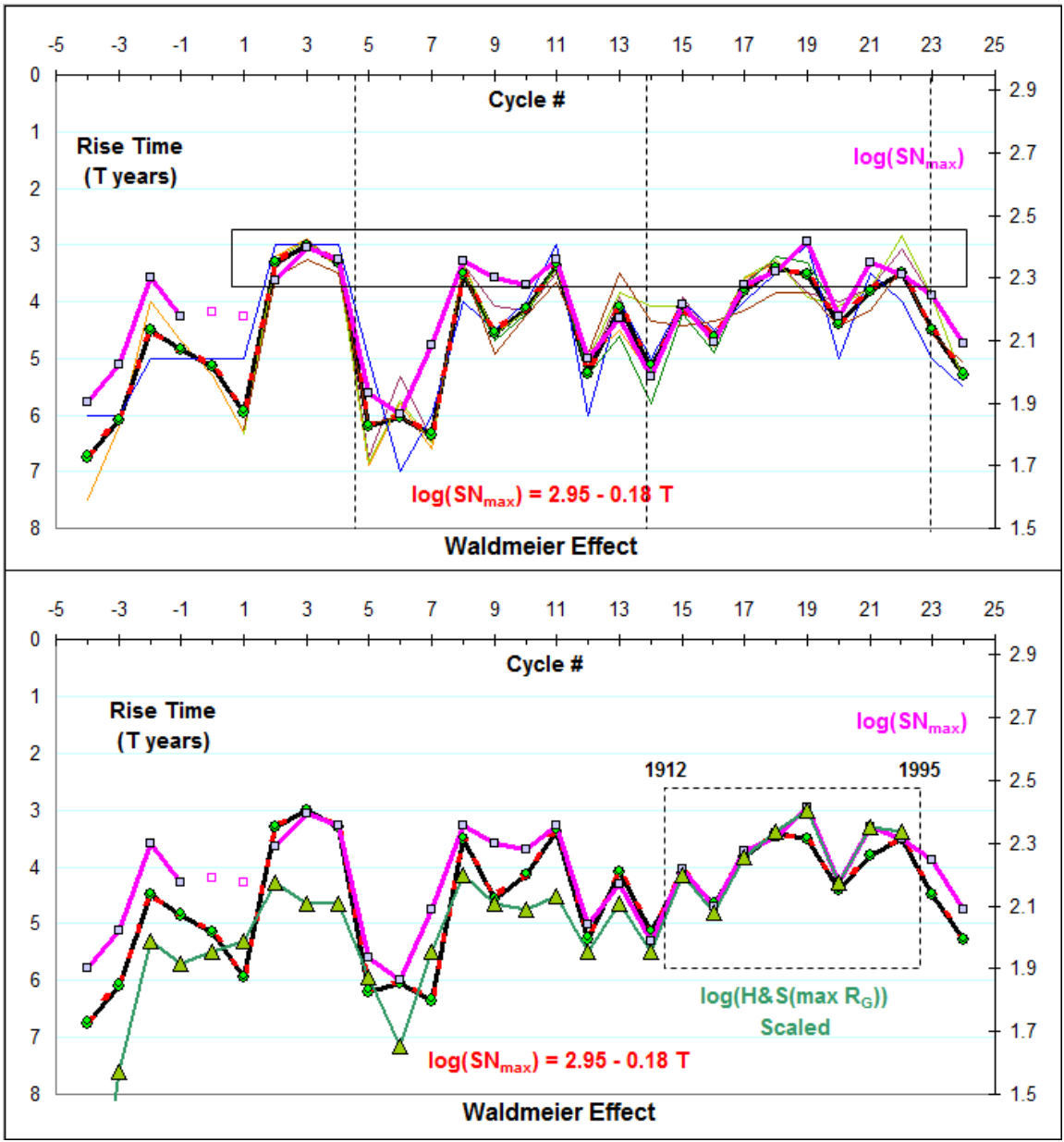

Figure 4: Upper panel: the logarithm of the maximum sunspot number (pink heavy curve with square symbols) and the rise time (heavy black curve) for cycles -4 through 24. The thin curves are for individual sources of the rise time. There are very few actual observations for cycles 0 and 1 (open pink squares). In the lower panel we show as a curiosum that the now defunct Hoyt and Schatten (1998) Group Sunspot Number (green curve with triangle symbols) does not exhibit the same stable Waldmeier Effect before Cycle 12, as also found by others (e.g. Ogurtsov and Jungner, 2012; Hathaway, 2015). The maximum

Group Sunspot Numbers have been scaled to fit the SN (version 2) during 1912-1995.

The box in Figure 4 highlights that in each century solar activity reached the same level, consistent with the similar rise times of the cycles. Ignoring the danger of over-interpreting noisy data, there may be indications that the sunspot numbers for Cycles -4 to 0 are a bit too high.

## 3. Filtering to Remove Short-period Fluctuations

Suitable filters should have Gaussian shapes in the frequency domain and effectively remove high frequency variations (Hathaway et al., 1999). A tapered (to make the filter weights and their first derivatives vanish at the end points) Gaussian filter, as used by Hathaway (2015), is given by

$$W(t) = \exp(-u) - (3-u)/\exp(2)$$

where $u(t) = (t/a)^2/2$ and $-2a+1 \leq t \leq 2a-1$

and $t$ is time from the center of the filter in a suitable unit (e.g. 1 month or 1 solar rotation) and $2a$ is the Full-Width-at-Half-Maximum of the filter in the same unit. The significant variations in solar activity on time scales of one to three years which can produce double peaked maxima are filtered out by a 24-month Gaussian filter, i.e. with $a = 12$ months. To remove the (large) rotational variation we also use 27-day Bartels rotation averages as well as the more traditional monthly means. To determine the times of minima and maxima we fit a quadratic to the smoothed data for a few years around the approximate minima and maxima and calculate the times of the extrema as shown in Figure 5 and listed in Table 2.

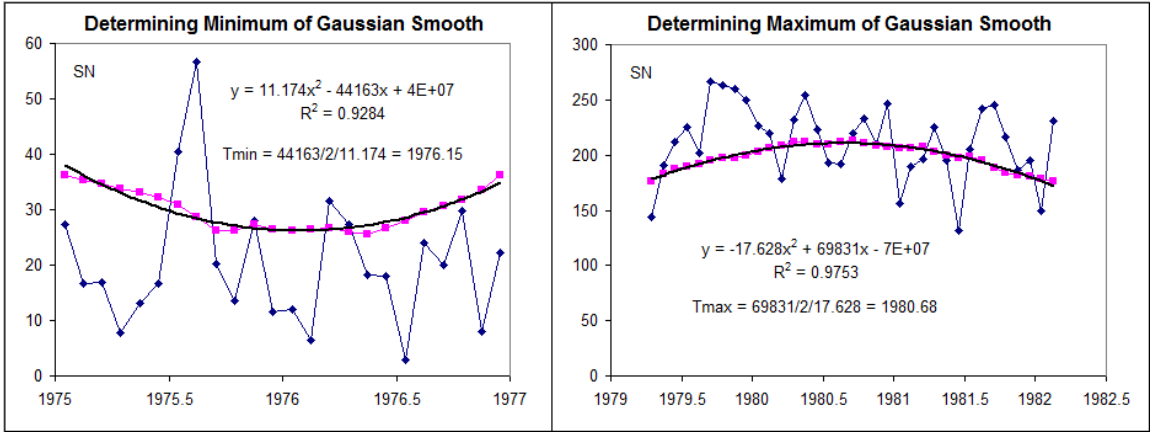

Figure 5: Left: Monthly average sunspot number (blue curve with diamonds) and Gaussian smoothed value (pink curve with squares). A second order polynomial fit (black curve) is used to determine the time of minimum. Right: Same, but for the maximum of Cycle 21.

The Bartels rotation data was constructed as described in Svalgaard (2016a) combining the yearly Group Number for the long-term variation with the daily Sunspot Number for intra-year variation (http://www.leif.org/research/WN-BROT.txt).

Table 2: Times of minima, maxima, and rise of Cycles 1 through 24 determined from the Gaussian smoothed values of monthly means of the sunspot numbers (left) and 27-day Bartels rotations.

| Cycle # | Monthly SN v2 | | | | Bartels Rotation WN | | | |
|---|---|---|---|---|---|---|---|---|
| | Min | Max | Rise T | log SN | Min | Max | Rise T | log SN |
| 1 | 1755.58 | 1761.39 | 5.80 | 2.069 | | | | |
| 2 | 1766.15 | 1770.37 | 4.22 | 2.223 | | | | |
| 3 | 1775.55 | 1779.01 | 3.46 | 2.338 | | | | |
| 4 | 1784.37 | 1788.57 | 4.19 | 2.327 | | | | |
| 5 | 1798.65 | 1804.07 | 5.43 | 1.841 | | | | |
| 6 | 1810.80 | 1816.90 | 6.10 | 1.856 | | | | |
| 7 | 1823.10 | 1829.63 | 6.54 | 2.034 | 1823.05 | 1829.71 | 6.66 | 2.068 |
| 8 | 1833.70 | 1837.56 | 3.86 | 2.315 | 1833.82 | 1837.37 | 3.55 | 2.333 |
| 9 | 1843.66 | 1848.84 | 5.18 | 2.282 | 1843.71 | 1848.86 | 5.15 | 2.268 |
| 10 | 1856.00 | 1860.43 | 4.43 | 2.232 | 1856.06 | 1860.39 | 4.33 | 2.265 |
| 11 | 1867.03 | 1871.20 | 4.18 | 2.296 | 1867.09 | 1871.17 | 4.08 | 2.333 |
| 12 | 1878.42 | 1883.67 | 5.25 | 2.015 | 1878.55 | 1883.82 | 5.27 | 2.049 |
| 13 | 1889.30 | 1894.04 | 4.74 | 2.119 | 1889.38 | 1893.77 | 4.38 | 2.152 |
| 14 | 1901.52 | 1906.78 | 5.26 | 1.994 | 1901.66 | 1906.72 | 5.06 | 1.970 |
| 15 | 1912.90 | 1917.97 | 5.08 | 2.153 | 1912.91 | 1918.05 | 5.14 | 2.140 |
| 16 | 1923.37 | 1927.95 | 4.58 | 2.076 | 1923.47 | 1928.03 | 4.56 | 2.078 |
| 17 | 1933.49 | 1938.25 | 4.76 | 2.247 | 1933.57 | 1938.16 | 4.59 | 2.252 |
| 18 | 1943.98 | 1948.37 | 4.39 | 2.299 | 1944.01 | 1948.27 | 4.27 | 2.287 |
| 19 | 1954.03 | 1958.36 | 4.33 | 2.414 | 1954.09 | 1958.33 | 4.24 | 2.394 |
| 20 | 1964.64 | 1969.31 | 4.67 | 2.168 | 1964.84 | 1969.31 | 4.46 | 2.171 |
| 21 | 1976.15 | 1980.68 | 4.53 | 2.327 | 1976.19 | 1980.52 | 4.34 | 2.330 |
| 22 | 1986.18 | 1990.54 | 4.36 | 2.306 | 1986.19 | 1990.57 | 4.38 | 2.296 |
| 23 | 1996.26 | 2001.30 | 5.04 | 2.230 | 1996.34 | 2001.33 | 4.99 | 2.250 |
| 24 | 2008.76 | *2014.20* | *5.44* | *2.050* | 2008.71 | *2013.90* | *5.19* | *2.060* |

In Figure 6 we show the result for both the monthly and Bartels rotation averages, again confirming the stability of the sunspot number calibration.

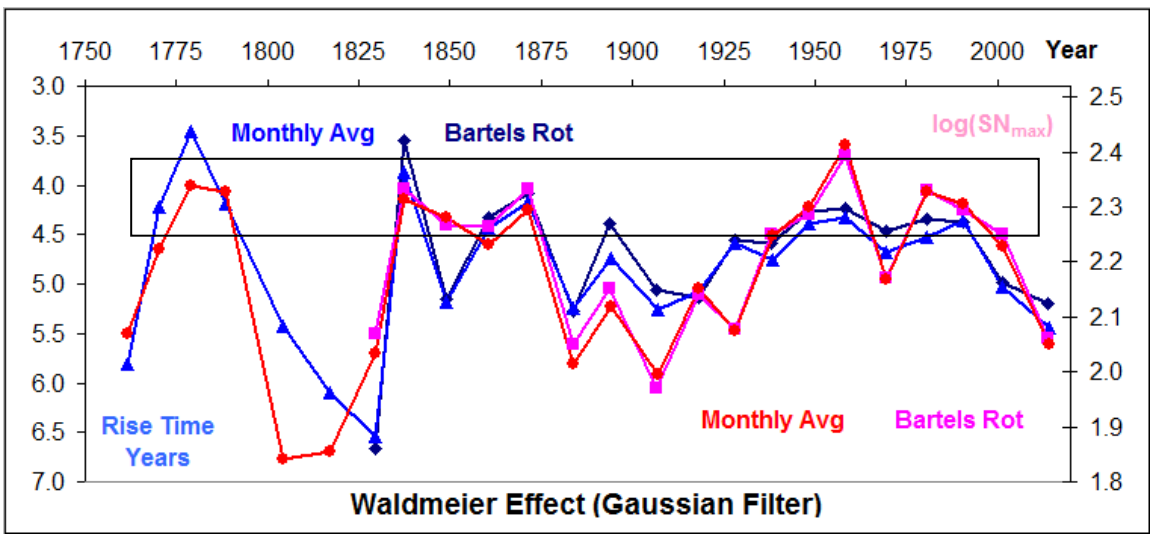

Figure 6: Rise times (bluish curves with triangle and diamond symbols) and cycle amplitudes (reddish curves with square and circle symbols) based on Gaussian filtering over 2-year (24 months or 27 Bartels rotations) FWHM windows.

## 4 Waldmeier Effect also Found in Sunspot Areas

To detect the Waldmeier Effect for sunspot areas we use a composite sunspot area series (cross-calibration of measurements by different observatories of observed values not corrected for projection, just as the sunspot numbers) maintained by Balmaceda et al. (2009, http://www2.mps.mpg.de/projects/sun-climate/data/table4_v0613.txt) with up-to-date values through 2015 from Hathaway scaled to match Balmaceda's since 2009 (http://solarscience.msfc.nasa.gov/greenwch/daily_area.txt). Monthly average areas, SA, are first transformed to equivalent sunspot numbers using $SA^* = 0.5073\, SA^{0.732}$ (see e.g. Clette et al., 2014, Figure 37). Using the same Gaussian 24-month filter we determine the values given in Table x and shown in Figure 7. It is evident that the Waldmeier Effect is also present for sunspot areas (Figure 7 at right) in general agreement with the Effect in sunspot numbers.

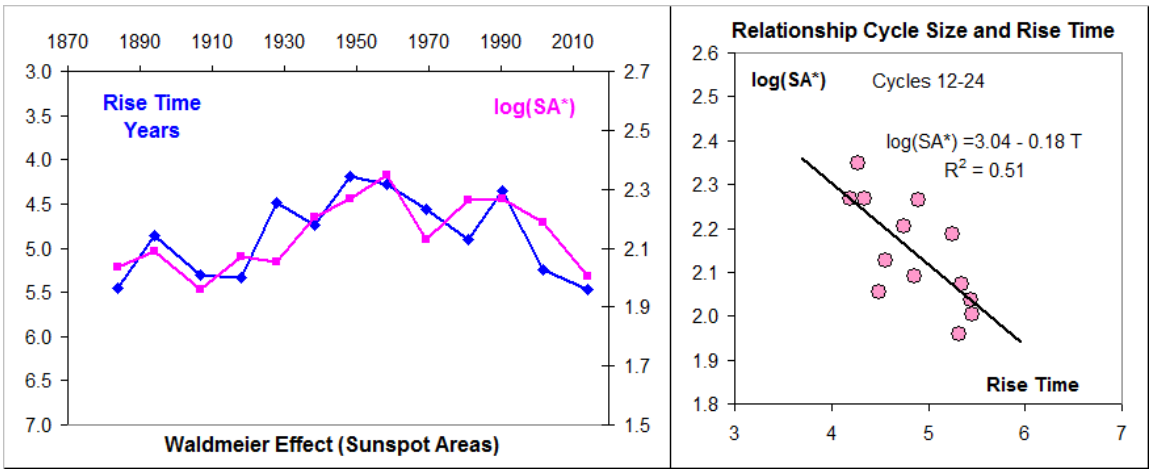

Figure 7: Waldmeier Effect for equivalent sunspot numbers derived from observed sunspot areas (pink with square symbols) using the 24-month Gaussian filter and the procedure explained in Figure 5.

Table 3: Times of minima, maxima, and rise of Cycles 1 through 24 determined from the Gaussian smoothed values of monthly means of the sunspot number equivalent derived from the observed sunspot areas.

**Sunspot Areas Showing Waldmeier Effect**

| Cycle # | Min | Max | Rise Time | log(SA*) |
|---|---|---|---|---|
| 12 | 1878.40 | 1883.85 | 5.45 | 2.036 |
| 13 | 1889.06 | 1893.92 | 4.86 | 2.091 |
| 14 | 1901.41 | 1906.72 | 5.31 | 1.958 |
| 15 | 1912.89 | 1918.23 | 5.34 | 2.072 |
| 16 | 1923.43 | 1927.93 | 4.49 | 2.054 |
| 17 | 1933.55 | 1938.30 | 4.74 | 2.204 |
| 18 | 1944.00 | 1948.20 | 4.19 | 2.266 |
| 19 | 1954.09 | 1958.37 | 4.29 | 2.347 |
| 20 | 1964.70 | 1969.25 | 4.56 | 2.128 |
| 21 | 1976.10 | 1980.99 | 4.90 | 2.265 |
| 22 | 1986.34 | 1990.69 | 4.35 | 2.266 |
| 23 | 1996.42 | 2001.67 | 5.25 | 2.187 |
| 24 | 2008.88 | *2014.34* | *5.46* | *2.004* |

The question "The Waldmeier Effect: An Artifact of the Definition of Wolf Sunspot Number?" raised by Dikpati et al. (2008) based on their analysis of sunspot areas can thus be answered in the negative: the Waldmeier effect is a very real solar effect. A conclusion also reached by Karak and Choudhuri (2011).

## 5 Waldmeier Effect also Found for EUV and F10.7 Proxies

Solar EUV radiation (with wavelength less than 103 nm) creates and maintains the E-layer of the ionosphere. The resulting conductive medium moves across the Earth's magnetic field maintaining a dynamo current whose magnetic effect we observe at the surface. The effect is manifested mainly in the East-component of the geomagnetic field as a diurnal variation whose magnitude (*rY*) is a measure of the EUV flux (and of its proxy, the F10.7 microwave flux) and thus shows a clear sunspot cycle variation, Figure 8 (Svalgaard, 2013, 2016b), that can be used to derive the Waldmeier Effect, Figure 9.

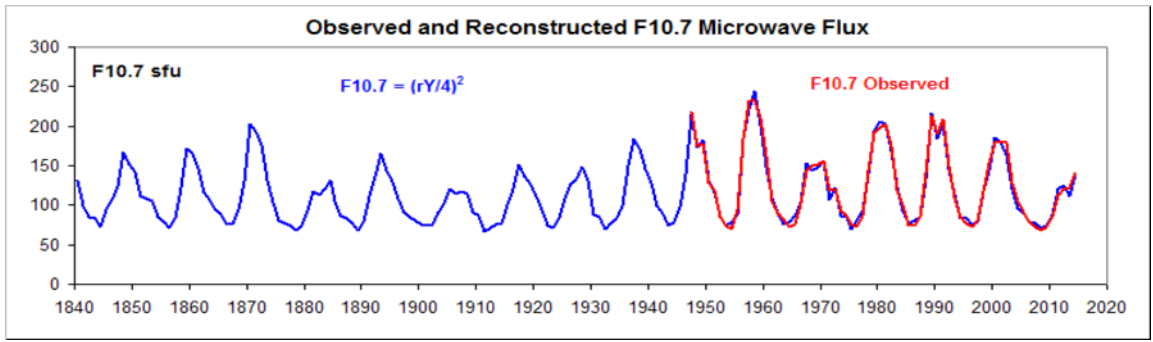

Figure 8: Yearly average values of the F10.7 microwave flux (for 1947-2014, red) compared to values reconstructed from the range, *rY*, of the diurnal variation of the East-component of the geomagnetic field (for 1840-2014, blue).

Even before the 'Magnetic Crusade' of the 1840s we have scattered observations of the diurnal variation of the Declination. Figure 10 shows the early data mainly collected and published by Wolf and reduced by Loomis (1870) to the common scale of Prague. The Waldmeier Effect is also visible in these early data as well as the general agreement between the Group Number and the geomagnetic proxy supporting our conclusion about the stability of the scale values.

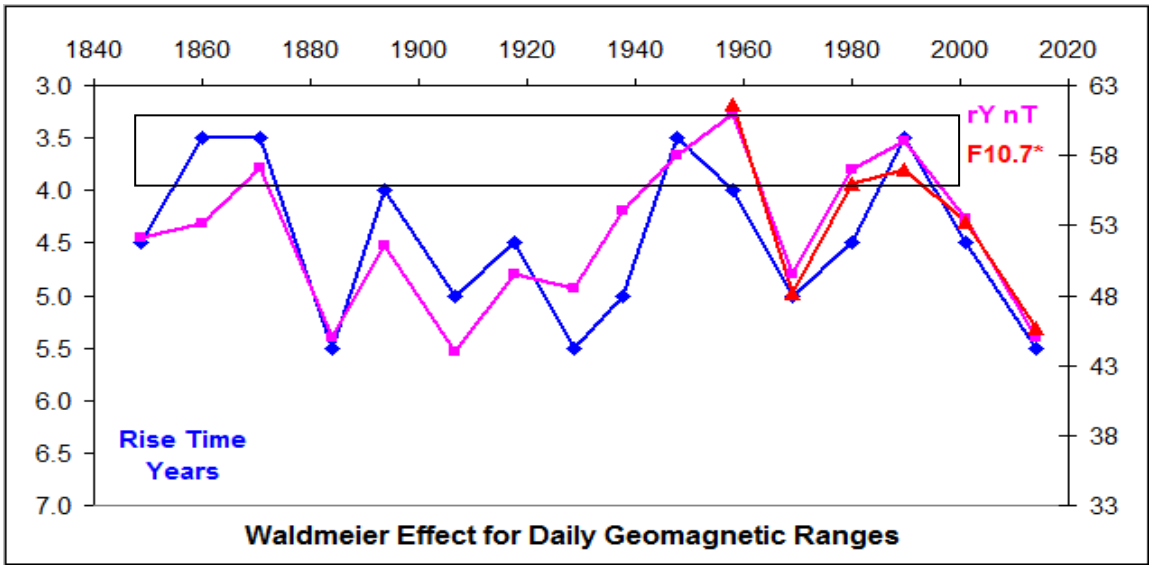

Figure 9: Waldmeier Effect derived from the range (pink), *rY*, of the diurnal variation of the East-component of the geomagnetic field and from the F10.7 flux (red). The rise times (blue) determined from yearly values to a resolution of half a year.

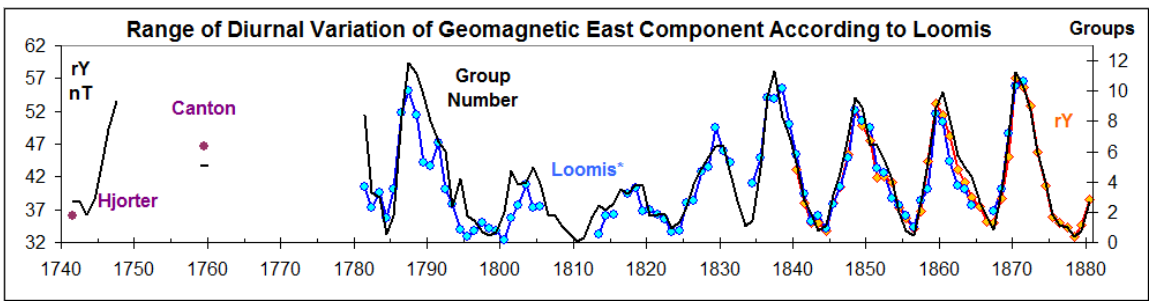

Figure 10: The range, *rY*, of the diurnal variation of the geomagnetic East component determined from the daily range of Declination (given by Loomis, 1870) converted to force units in the East direction (blue curve) and then scaled to match *rY* (Svalgaard, 2016b, red curve), supplemented with observations by Canton, 1759, and Hjorter, 1747. The Sunspot Group Number (Svalgaard and Schatten, 2016) is shown (black curve without symbols) for comparison scaled (right-hand scale) to match *rY*.

Loomis drew two important and prescient conclusions: 1) the basal part of the "diurnal inequality (read: variation), amounting at Prague to six minutes is independent of the changes in the sun's surface from year to year", and 2) "the excess of the diurnal inequality above six minutes as observed at Prague, is almost exactly proportional to the amount of spotted surface upon the sun, and may therefore be inferred to be produced by this disturbance of the sun's surface" (and may thus serve as an independent measure of solar activity). It is encouraging that the new Sunspot Group Number series seems to agree well with diurnal range series, even for the earliest geomagnetic measurements. Loomis' conclusions are fully supported by our modern data and analyses.

## 6. The 'Other' Waldmeier Effect

If we define the 'growth rate', $g$, of a cycle as its maximum sunspot number divided by the rise time, the 'normal' Waldmeier Effect implies that $g = SN_{max}/T$ also should be larger for large cycles than for small cycles, and so it is, Figure 11.

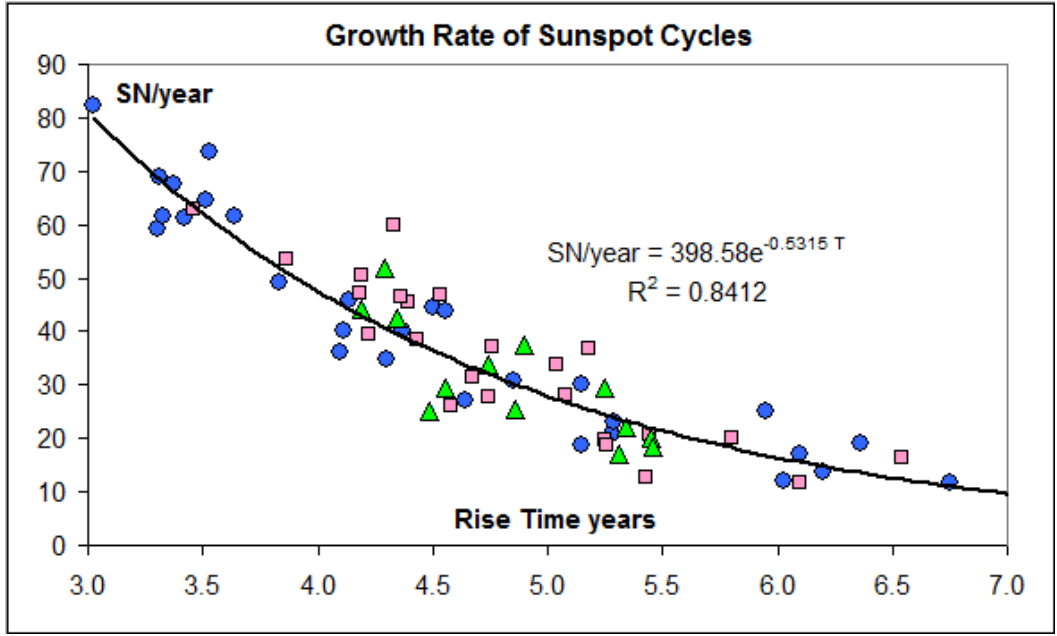

Figure 11: The growth rate (Sunspot Numbers per year) as a function of the rise time as derived from the data in Table 1 (blue circles), Table 2 (pink squares), and Table 3 (green triangles).

Karak and Choudhuri (2011), Roshchina and Sarychev (2014), and Ogurtsov (2012) further discuss this second Waldmeier Effect, albeit with slightly different definitions of how to measure the growth rate. From the fit to the data points in Figure 11 we can recover yet another formula for the 'first' Waldmeier Effect for cycle $n$: $SN_{max}(n) = g(n) T(n)$ or (approx. with a suitable calibration constant $a$) $SN_{max} = aT / \exp(T/2)$.

## 7. Historical Note

Although Max Waldmeier today is credited with "the Waldmeier Effect" for the finding that large sunspot cycles have shorter rise times than do small cycles, this fact was known already to Wolf (we are still basically using his determinations of the times of early minima and maxima) and was seriously discussed around the turn of the 20[th] century (e.g. Halm 1901, 1902; Lockyer 1901; and Wolfer 1902) and taken as evidence for an 'eruption-type' sunspot cycle freed from "the shackles of unduly close adherence to harmonic analysis" (Milne 1935). Figure 12 shows a "quite remarkable" correlation between the rise time and the size of the cycle, so "that the interval between the minima and maxima depends intimately upon the intensity of development of the spots during that period would seem therefore to be quite certain" (Wolfer 1902).

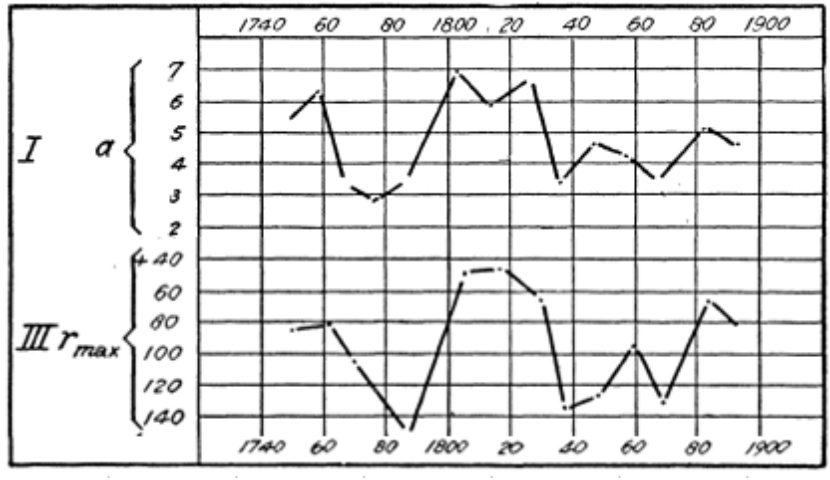

Figure 12: I: The rise time (in years) for each cycle 1750-1890. III: The maximum smoothed relative sunspot numbers for each cycle (adapted after Wolfer, 1902). Note, that the scales are reversed compared to Figure 4.

## 8. Why is the Solar Cycle Asymmetric?

Kitiashvili and Kosovichev (2011) show that the Waldmeier Effect can have an explanation in terms of non-linear Dynamo Theory augmented by the idea of conservation of magnetic helicity. Because of conservation of the total helicity, a growth of the large-scale magnetic helicity due to the dynamo action is compensated by the growth of the small-scale helicity of opposite sign, leading to asymmetries of the solar cycle. The evolution of magnetic helicity and its nonlinear feedback on the generation of the large-scale magnetic field by the α-effect as advocated by Pipin and Kosovichev (2011) also lead to asymmetric sunspot cycles. Karak and Choudhuri (2011) find that a flux transport solar dynamo with high turbulent diffusivity is consistent with the Waldmeier Effect while models with low diffusivity lead to the opposite of the effect. So, there is no shortage of 'understanding' of possible physical causes of the Waldmeier Effect. In any event, the now well-established Waldmeier Effect is a firm observational constraint that any theory of the solar cycle must explain.

## 9. Conclusion

We show that the Waldmeier Effect derived in different ways from several updated manifestations of the solar activity record (spots, groups, areas, microwave flux, and terrestrial proxies) is robust and confirms Waldmeier's earlier conclusion that the 'scale values' of the 'official' sunspot records have been constant or, at least, have undergone no significant changes over the last (now) 250 years The implication is that solar activity reached similar high values in every one of the 18$^{th}$, 19$^{th}$, and 20$^{th}$ (and perhaps early in the 17$^{th}$) centuries, supporting the mounting evidence that there has been no modern *Grand* Maximum. That the Hoyt and Schatten (1998) Group Sunspot Number does not exhibit a stable Waldmeier Effect has been resolved and can be understood from the recent reconciliation of the sunspot and group records. We therefore urge the community to fully embrace the newly revised series.

## Acknowledgements

"There is nothing more difficult to take in hand, more perilous to conduct, or more uncertain in its success, than to take the lead in the introduction of a new order of things" (Niccolò Machiavelli, The Prince). We thank Irina Kitiashvili for comments. LS thanks Stanford University for continuing support.